\documentclass[11pt]{article}

\usepackage[a4paper,margin=25mm]{geometry}
\usepackage{amsmath,amssymb,bm}
\usepackage{graphicx}
\usepackage{booktabs}
\usepackage{tabularx}
\usepackage{microtype}
\usepackage[numbers,sort&compress]{natbib}
\usepackage{bibunits}
\usepackage[hidelinks]{hyperref}
\usepackage[mathlines]{lineno}
\modulolinenumbers[1]

\newcommand{\iu}{\mathrm{i}}
\newcommand{\Tr}{\operatorname{Tr}}

\title{\textbf{Quantum--classical break-even \\in electronic dynamics of macrocyclic molecules}}

\author{\parbox{0.94\textwidth}{\centering
Shu Kanno$^{1,2*}$, Kenji Sugisaki$^{3,2,4}$, Takashi Imamichi$^{5}$, 
Toshinari Itoko$^{5,2}$,\\
Rei Sakuma$^{6,2}$,
Hajime Nakamura$^{2}$, Qi Gao$^{1,2}$ and Naoki Yamamoto$^{2,7}$\\[0.6em]
\small $^{1}$Mitsubishi Chemical Corporation, Science \& Innovation Center, Yokohama 227-8502, Japan\\
\small $^{2}$Quantum Computing Center, Keio University, 3-14-1 Hiyoshi, Kohoku-ku, Yokohama 223-8522, Japan\\
\small $^{3}$Deloitte Tohmatsu LLC, 3-2-3 Marunouchi, Chiyoda-ku, Tokyo 100-8363, Japan\\
\small $^{4}$Centre for Quantum Engineering, Research and Education, TCG Centres for Research and Education in Science and Technology, Sector V, Salt Lake, Kolkata 700091, India\\
\small $^{5}$IBM Research, 19-21 Nihonbashi Hakozaki-cho, Chuo-ku, Tokyo 103-8510, Japan\\
\small $^{6}$Materials Informatics Initiative, RD Technology \& Digital Transformation Center, JSR Corporation, 3-103-9 Tonomachi, Kawasaki-ku, Kawasaki 210-0821, Japan\\
\small $^{7}$Department of Applied Physics and Physico-Informatics, Keio University, 3-14-1 Hiyoshi, Kohoku-ku, Yokohama 223-8522, Japan}}
\date{}
\makeatletter
\let\savedmaketitle\maketitle
\let\savedatmaketitle\@maketitle
\let\savedtitle\title
\let\savedauthor\author
\let\saveddate\date
\makeatother

\begin{document}
\begin{bibunit}[naturemag]
\maketitle
$^{*}$Correspondence: \texttt{shu.kanno@quantum.keio.ac.jp}

\begin{abstract}
Here we demonstrate quantum--classical break-even for chemical dynamics applications, reaching useful accuracy with a quantum-hardware wall-clock time comparable to that estimated for practical classical computing resources. 
To this end, we develop a workflow for problems that are hardware-native but classically hard. We target large ring-shaped molecules that map naturally onto quantum hardware with neighboring connections. Exploiting the spatial locality of interactions, we compress short-time evolution within a small region using tensor networks. We then use rotational symmetry to replicate the compressed circuit in space and time, enabling long-time dynamics of large systems with practical classical processing times.
We apply the workflow to conjugated macrocycles with rotational periodicity, 12-qubit benzene and 120-qubit [60]annulene, by embedding their electronic structure onto loops of IBM quantum processors. For the annulene, we compared the quantum-processing-unit (QPU) wall-clock times for hole-doped nonequilibrium dynamics to the estimated wall-clock time of two state-of-the-art classical methods, Majorana propagation and matrix-product-state time evolution based on the time-dependent variational principle.  
We show a situation in which a hybrid quantum--classical strategy is comparable to a classical-only counterpart in terms of both accuracy and computational time for a conventional high-performance computing system under ideal parallelization. These results show that quantum hardware can become a realistic computational option for chemical simulations for practitioners in chemical science and engineering.
\end{abstract}


\begin{figure}[t]
    \centering
    \includegraphics[width=0.9\linewidth]{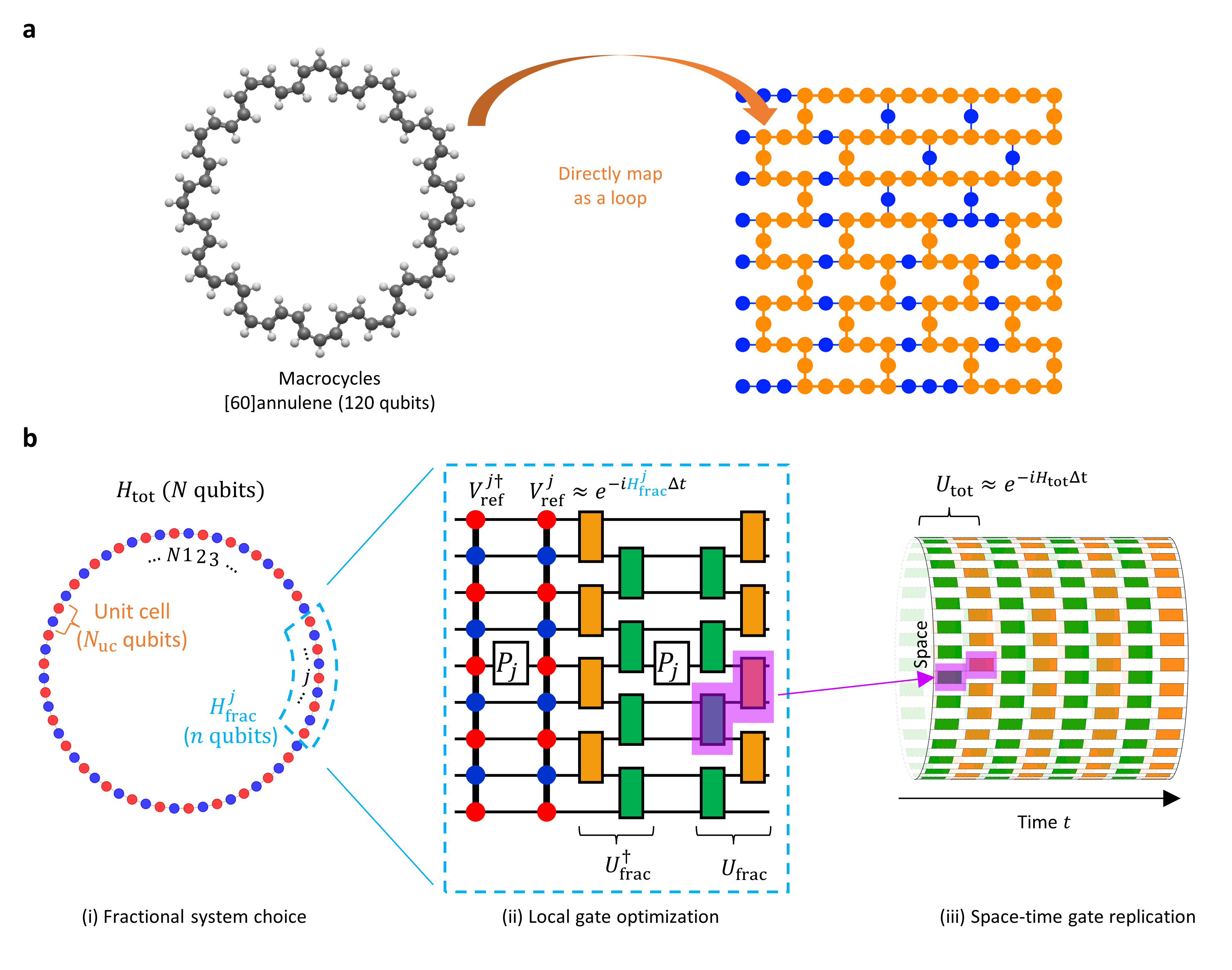}
    \caption{Overview of the quantum workflow for dynamics of cyclic molecules. (a)  [60]annulene (120-qubit) model and the mapped loop adopted in this study. (b) Workflow including the fractional system choice, local (translationally invariant) gate optimization, and space-time gate replication. This workflow constructs a quantum circuit to simulate a ring molecule by replicating a fundamental unit cell. This unit cell functions as a unitary gate designed to evolve the quantum state by a prescribed time step $\Delta t$, and its repeated execution achieves time evolution of the total system under $H_{\rm{tot}}$ for a total time $t$ that is an integer multiple of $\Delta t$.}
    \label{fig:workflow}
\end{figure}

Chemical calculations underpin a broad range of applications, including
semiconducting materials, conducting materials, catalysis and optoelectronic
materials~\cite{Van-de-Walle2004-oc,Bredas1982-ln,Norskov2009-fr,Hachmann2011-tb}. Electrons in chemical systems obey quantum mechanics, and the
Hilbert space required to represent their many-electron states grows
exponentially with system size. Consequently, accurate electronic-structure
calculations rapidly become computationally demanding as the system size
increases. Quantum computers provide a complementary route by directly
manipulating many-body quantum states in an exponentially growing Hilbert
space, rather than explicitly representing all of their amplitudes on a
classical computer.

A central objective in applying quantum algorithms is to perform quantum computation in less time than, or in a time competitive with, state-of-the-art classical computation to solve a certain task. 
Hereafter we call the former and latter quantum advantage and quantum break-even, respectively. They have been claimed in quantum dynamics of physical lattice models, including Ising models and Hubbard models~\cite{King2025-rq, Zheng2026-sy, Hartnett2026-cq},
while the advantage can be revised by the development of sophisticated classical
algorithms~\cite{Ouyang2026-ni, Tindall2026-qg}. 

Nevertheless, reaching such break-even for chemical
models is particularly challenging, although quantum utility has been claimed under the less stringent criterion of performing reliable computations beyond the reach of brute-force classical computation~\cite{Kim2023-xg, Robledo-Moreno2025-ul}.
Current quantum processors have inevitable physical noise and thus limitations on executable circuit depth. In addition, device-dependent operational constraints exist, such as fixed device topologies in solid-state devices and particle-transport overheads in particle-trap devices~\cite{Kim2023-xg, Brown2016-be}. 
Chemical Hamiltonians are
determined by their electronic structure and generally contain more complex
interactions that cannot be freely redesigned to suit the device.
To achieve the break-even, strategies that are not only classically hard but also quantumly easy should be explored.

We focus on macrocycles in Fig.~\ref{fig:workflow}a as a matched target in this strategy. Macrocycles are molecules containing a large
ring-shaped framework and are widely used in areas such as photonic and electronic materials, pharmaceuticals, and molecular sensors~\cite{Lou2023-gn}. 
Also, a recent quantum computing work on a half-M\"obius molecular topology shows unusual
electronic structures~\cite{Roncevic2026-kk}.
On the classical side, in addition to complex interactions, they have a periodic structure, which makes their dynamics difficult to simulate with some strong classical algorithms. For example, tensor-network methods based on
matrix product states are particularly powerful for open
one-dimensional systems, whereas quantum dynamics in cyclic geometries are more
demanding~\cite{Schollwock2011-im}. 

On the quantum side, the cyclic
structure of a certain macrocycle can be mapped directly onto a closed loop in the
connectivity graph of a superconducting quantum processor (Fig.~\ref{fig:workflow}a),
avoiding long-range qubit operations. In addition, circuit compression
provides a further route to efficient execution by approximating the short-time
evolution generated by a complex chemical Hamiltonian with a shallow,
hardware-efficient circuit. 

Our workflow is illustrated in Fig.~\ref{fig:workflow}b, which has three steps: (i) fractional system choice in the left panel, (ii) local gate optimization in the middle panel, and (iii) space-time gate replication in the right panel. 
In this study, we assume periodic systems that have unit cells since they can save much of the classical cost for the circuit compression, where the number of qubits per unit cell is denoted as $N_{\mathrm{uc}}$ (two in the left panel).
For step (i), starting from the
full-system Hamiltonian $H_{\mathrm{tot}}$, we select a fractional Hamiltonian around the $j$-th qubit 
$H_{\mathrm{frac}}^{j}$ containing the local interactions relevant to a small
spatial region of $n$ qubits (nine in the panel), where $H_{\mathrm{tot}} = H_{\mathrm{frac}}^{j} + H_{\mathrm{res}}^{j}$ and $H_{\mathrm{res}}^{j}$ is a residual Hamiltonian. 
Note that theoretically, the value of $n$ can be determined by interaction length; technically, by the causal cone~\cite{Mizuta2022-ql}.

For step (ii), a shallow short-time circuit is
optimized classically within this fractional system by minimizing an overlap cost defined using a local Hilbert--Schmidt test (LHST):
\begin{equation}
C_{\mathrm{LHST}}^{\mathrm{per}}
=\frac{1}{N_{\mathrm{uc}}}
\sum_{j\in\mathcal S_{\mathrm{uc}}}C_{\mathrm{LHST}}^{j},
\label{eq:s_lhst_periodic}
\end{equation}
with
\begin{equation}
C_{\mathrm{LHST}}^{j}
=1-\frac14\left[
1+\frac{1}{2^n}\sum_{P_j\in\{X_j,Y_j,Z_j\}}
\Tr\!\left(V_{\rm{ref}}^{j\dagger} P_j V_{\rm{ref}}^j U_{\rm{frac}}^\dagger P_j U_{\rm{frac}}\right)
\right],
\label{eq:s_lhst_trace}
\end{equation}
where $\mathcal S_{\mathrm{uc}}$ is the set of qubit indices in one unit cell, $V_{\rm{ref}}^{j} \approx e^{{-i H_{\mathrm{frac}}^{j}\Delta t}}$ is a reference matrix product operator for accurate time evolution, $U_{\rm{frac}}$ is a circuit to be optimized having brick-wall gates with translational invariance under shifts by one unit cell (the depth $d$ is two in the panel), and $P_j$ is the Pauli gate of the $j$-th qubit. 
Note that the LHST can be defined for systems without rotational symmetry by changing the range of $j$ from one unit cell to the full system; see Supplementary Note~1 for details. Also, the extension to higher-dimensional systems has been proposed~\cite{Mizuta2022-ql}.

Although the common cost functions for circuit optimization using tensor networks are HST~\cite{Gibbs2025-hd, Le2025-ne} or the Frobenius norm~\cite{Mc_Keever2023-bu, Causer2024-wd, Mc_Keever2024-ji, Karacan2026-zh, Kanno2025-kj, Kanno2026-kv}, LHST enables the use of fractional systems, which avoids the limitation of dimensionality for tensor-network-based optimizations~\cite{Mizuta2022-ql, D-Anna2025-wk, Karacan2026-vf}. 
In addition, \(C_{\mathrm{LHST}}\) is defined in a \(2n\)-qubit Hilbert space using Bell measurements~\cite{Khatri2019-wb, Mizuta2022-ql}, but by expanding the Bell-measurement projectors in the Pauli basis and exploiting channel--state duality, \(C_{\mathrm{LHST}}\) can be formulated using only \(n\) qubits. These techniques enable more efficient circuit optimization using tensor networks than the conventional formulation.
The detailed
definition and tensor-network implementation of this objective are given in Supplementary Note~1. 

The same optimized local structure can represent symmetry-equivalent regions for a periodic system.
Thus, for step (iii), the resulting gates are replicated around the molecular ring and then along
the time direction to construct the full long-time circuit executed on the
quantum processor. We can calculate non-equilibrium and symmetry-breaking dynamics by setting a proper initial state.
The nonequilibrium hole dynamics considered in this study, evolving on a timescale of tens of attoseconds, may be relevant to ultrafast electronic dynamics following hole creation by extreme-ultraviolet or X-ray ionization~\cite{Krausz2009-jf}.
Thus, only the small-space, short-time optimization is
performed classically, whereas the resulting large-system, long-time dynamics
are assigned to quantum hardware.

We compare the computational time and accuracy of the quantum computation with those of state-of-the-art classical methods, the Majorana-propagation~\cite{Miller2025-bm}, and the time-dependent variational principle (TDVP) for matrix-product states (MPS)~\cite{Haegeman2011-ug}. We further discuss scenarios in which a hybrid strategy of quantum computation and Majorana propagation could achieve break-even.

We mention the scope of this study.
We primarily examine the regime in which the quantum-processing-unit (QPU) execution time is approximately 100 times shorter than the single-node runtime of the classical method required for the dynamics. We chose this regime because conventional high-performance computing (HPC) systems typically comprise hundreds to thousands of nodes~\cite{Shu2025-jm}, making this a reasonable estimate of the break-even boundary for users performing chemical calculations when ignoring the classical overhead associated with inter-node parallelization.
We also mention that the workflow includes not only the QPU time but also classical processing times. For the annulene model, the calculation times for the workflow were on the order of hours for Hamiltonian construction, days for circuit compression, seconds for preparing execution such as transpilation, and minutes for execution on the quantum hardware, excluding waiting time.
The quantum-hardware execution time includes not only the QPU execution time but also associated pre- and post-processing, such as pulse scheduling and error mitigation, and we use the QPU time as a main criterion when comparing the computational time with that of classical calculations, while we also show the total duration of the step (excluding queue-waiting time) for reference. Although these other times, especially the compression time, should in principle be considered, we exclude them here because a compressed circuit can be reused, e.g., when changing initial states and inserting local perturbations on the circuit. Thereby, we can reduce their relative contribution to the total computational time for specific tasks.
The end-to-end time analysis remains a future research direction to achieve broad chemical applications.

\section*{Benzene validation on quantum hardware}
\label{sec:benzene_main}

We first validate the local compression procedure for a benzene $\pi$-electron model with six localized carbon $p_z$ orbitals, corresponding to 12 qubits. The model has $D_{\rm{6h}}$ symmetry, and $N_{\rm{uc}}=2$. We also use assumptions to impose short-range interactions and symmetries on the model, such as setting a cutoff interaction distance and avoiding long-range Pauli operators in a boundary term using a parity transformation. See the details of Hamiltonian construction in Methods. The Hamiltonian includes 951 Pauli terms. 
Using the interleaved ordering $q_1=(1\uparrow)$,
$q_2=(1\downarrow)$, $q_3=(2\uparrow)$, $q_4=(2\downarrow),\dots$, 
the initial computational-basis state is $\lvert q_{1} q_{2} \dots q_{12}\rangle  = \lvert100110011000\rangle$, giving
a hole-doped antiferromagnetic state. The target observable is the one-body Pauli $Z$ operator on the first qubit, $Z_1$, which corresponds to the electron number as $\hat{n}_1 = (1-Z_1)/2$ under the Jordan-Wigner transformation. The time points are $t=\{0.2, 0.4,\dots, 2.0\}$ with $\Delta t = 0.2$, both in atomic units (one atomic unit of time is about 24 attoseconds). The depth $d=5$, and the fractional cell size $n=9$.
We used quantum hardware ibm\_kawasaki; see the Methods section for details of Hardware settings.

\begin{figure}[]
    \centering
    \includegraphics[width=0.5\linewidth]{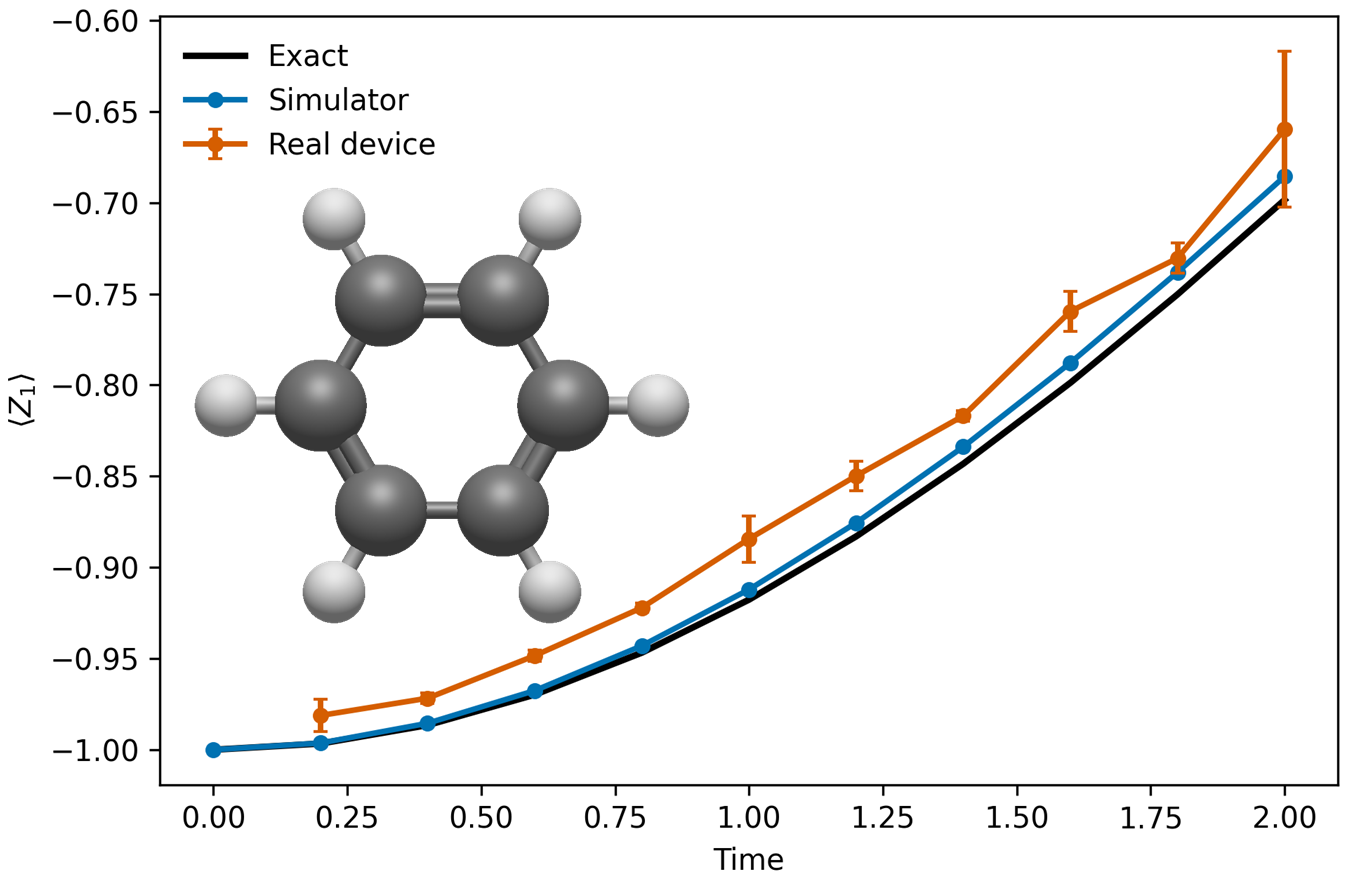}
    \caption{The expectation value calculation for benzene. The error bar comes from a standard deviation for zero-noise extrapolations.}
    \label{fig:benzene}
\end{figure}

The result is shown in Fig.~\ref{fig:benzene}. 
The spatially replicated circuit on a noiseless simulator (blue) closely reproduces the
exact dynamics (black) through $t=2.0$, with an overall mean/median absolute error of 0.0066/0.0065. 
The same circuit executed on quantum hardware also follows the exact dynamics, with a mean/median absolute error of 0.027/0.025.
The number of 2-qubit (controlled-Z) gates and the depth of the circuit at $t=2.0$ are 738 and 123, respectively, which are much smaller than those of the first-order Trotter circuit (without optimization) for the device, 256,240 and 172,526, respectively.
The results show that the workflow, including the gate optimization and replication, can be efficiently applied in practical situations.

\begin{figure}[h]
    \centering
    \includegraphics[width=1\linewidth]{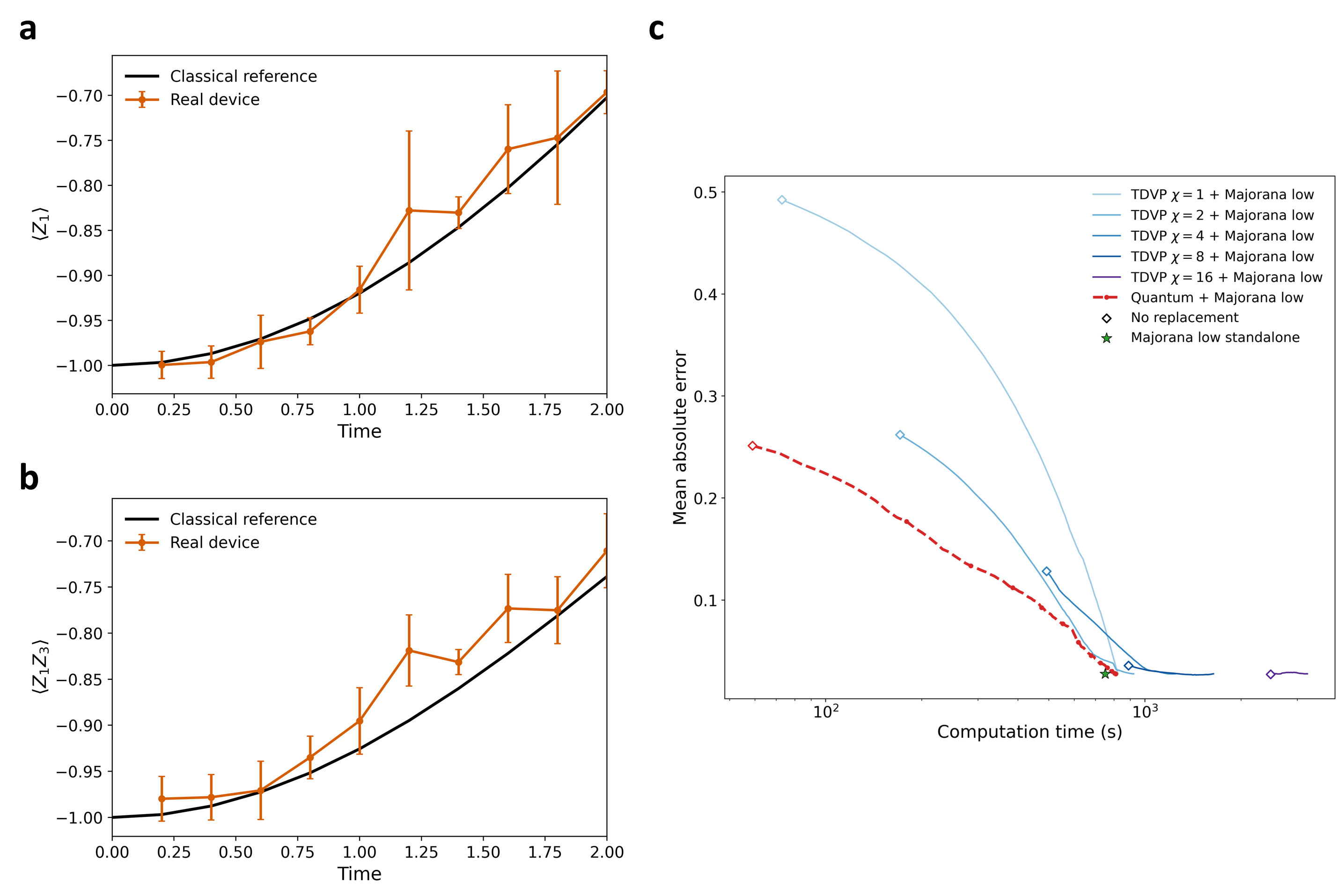}
    \caption{Electronic dynamics of the 120-qubit [60]annulene
    model. 
    (a) An example for $\langle Z_1 \rangle$. (b) That for $\langle Z_1 Z_3 \rangle$. (c) Time-to-accuracy trade-off at \(t=2.0\) for all 7,260 one- and two-qubit
\(Z\) observables. The horizontal axis assumes ideal parallel execution of the Majorana propagation (low accuracy)
over 100 nodes.
The circular markers on the dashed curve indicate increments of ten selected logical qubits.}
\label{fig:annulene_120q}
\end{figure}

\section*{Quantum--classical break-even in [60]annulene}
\label{sec:benchmark_main}

We next scale the workflow to a 120-qubit [60]annulene model and compare the
quantum calculation with a classical reference.
Below, we first provide an overview of the calculations, then present the results from the independent quantum and classical calculations, and finally discuss a break-even situation achieved using a hybrid quantum--classical strategy.

Large annulene-type cyclic polyacetylenes have been synthesized only as ensembles rather than as isolated, size-defined
$[M]$annulenes~\cite{Miao2021-hj}. 
Large annulenes are nevertheless theoretically
interesting because increasing the ring size can qualitatively change their
electronic properties: in particular, theoretical calculations have suggested
a progressive loss of H\"uckel and Baird aromaticity with increasing annulene
size, approaching essentially non-aromatic behaviour for sufficiently large
rings~\cite{Van-Nyvel2025-kt}. 
Also, from an application perspective, understanding electron dynamics in large-scale \(\pi\)-conjugated systems is useful for elucidating the mechanisms of charge transfer and excitation energy transfer in photochemical reactions and molecular electronics~\cite{Fratini2020-kv, Dimitriev2022-tx}.

The [60]annulene model contains 60 localized $\pi$ spatial orbitals, corresponding to 120 spin-orbitals and 120 qubits. 
The model was constrained to have $D_{\rm{12h}}$ symmetry and has five spatial orbitals per unit cell, $N_{\rm uc}=10$, and $n=17$; see the Methods section for details of Hamiltonian construction. The Hamiltonian includes 8,597 Pauli terms.
The initial state is a hole-doped antiferromagnetic state with fluctuations (see Hardware setting in Methods).
We measured not only one-body Pauli-$Z$ observables but also two-body ones, $ \{Z_l\}_{l=1}^{120}\cup\{Z_lZ_m\}_{1\leq l<m\leq120}$, which are related to two-body occupation correlations $\langle\hat n_l\hat n_m\rangle$. The number of observables becomes $N_{\rm obs}=\binom{120}{1}+\binom{120}{2}=7,260$. These observables commute with each other and can be evaluated without increasing QPU time. 
Since each observable was evaluated at
$t\in[0.2,2.0]$ with $\Delta t=0.2$, the total number of evaluated observables is 72,600. 
The compressed 120-qubit circuits were executed on ibm\_boston, and the qubit layout is shown in Fig.~\ref{fig:workflow}a. 

We adopt Majorana propagation in the \texttt{monoprop} package~\cite{Miller2025-bm} as one classical reference.
Majorana-string coefficients with magnitude below $\epsilon_{\mathrm{cut}}$
were discarded during propagation. We choose the high-accuracy setting of $\Delta t=0.05$ and $\epsilon_{\mathrm{cut}}=10^{-7}$ as the reference for observable values, and the low-accuracy setting of $\Delta t=0.2$ and $\epsilon_{\mathrm{cut}}=10^{-3}$ as the runtime reference.
We also choose TDVP for MPS in the \texttt{ITensors.jl} package~\cite{Fishman2022-et} as another classical reference. We choose $\Delta t=0.05$ because mean absolute errors of 0.1 or higher could not be eliminated at $t=2.0$ with $\Delta t=0.2$.
See the Methods for details of Hardware settings, Majorana-propagation dynamics, and Time-dependent variational principle for matrix-product states.
Note that in practice, the rotational symmetry of the time-evolution operator can be exploited to reduce the number of observables that need to be evaluated using the Majorana propagation. However, this symmetry can be broken easily without changing the number of two-qubit gates, for example by introducing random local operators. We therefore expect this effect to have only a limited impact on the break-even claims of this study.

Figure~\ref{fig:annulene_120q}a and b show the expectation values of \(Z_1\) and \(Z_1Z_3\), respectively. Both exhibit good agreement with the reference values and capture the diffusion of charge. 
The mean/median absolute errors for \(\langle Z_1\rangle\) and \(\langle Z_1Z_3\rangle\) are 0.016/0.008 and 0.026/0.022, respectively, with the errors being of the same order of magnitude as those obtained for benzene. 
The 2-qubit gate count and depth are 7,380 and 123, respectively. 

We first evaluated the results for all 7,260 observables. Table~\ref{tab:annulene_benchmark} summarizes the results obtained by applying the same analysis to all observables. For the error-mitigated results in the first row under the “All observables” group header, the mean/median absolute errors are 0.203/0.090, respectively, where the large mean error would come from specific low-accuracy qubits; see Supplementary Note 3 for low-fidelity qubit evaluation. 
We also report the QPU wall-clock time on the IBM Quantum Platform, which was 552~s, and the total batch-session wall-clock time including classical processing on the server, which was 1,456~s. Compared with the measured Majorana propagation (low-accuracy) wall-clock time of 75,113~s obtained on a single node with 36 CPU cores, these correspond to speedups of 136$\times$ and 52$\times$, respectively.

On the other hand, as shown in the last two rows of the table, all observables can be evaluated from a single MPS time evolution, whereas the computational time of TDVP increases rapidly with the bond dimension \(\chi\). Even relatively light calculations with \(\chi=8\) and \(16\) yield values for all observables with substantially higher accuracy on average than the quantum-hardware results; notably, previous hardware-validation studies have used MPS calculations with bond dimensions as large as \(\chi=4096\) as classical references~\cite{Hartnett2026-cq}. 
However, as shown under the “Boundary observable” group header, MPS methods do not efficiently handle periodic boundaries: for example, the accuracy of the boundary observable \(\langle Z_1 Z_{120}\rangle\) is comparable to that obtained on quantum hardware. 
The QPU time for \(t=2.0\) alone is 59~s, which is tens of times shorter than the corresponding TDVP runtime. Here, the TDVP results are based on the calculations for all observables, since the time required to evaluate the observables in TDVP is negligible compared with the time required for the time evolution.

\begin{table}[t]
    \centering

    \caption{Accuracy and runtime for the 120-qubit [60]annulene dynamics.
    In the first column, “Quantum” represents the time evolution results executed by the quantum device, and the others represent those by a classical computer.
    Mean and Median denote the mean and median absolute errors, respectively,
    over all observables and time points, relative to the high-accuracy Majorana-propagation results.
    The Boundary columns report the absolute error and runtime for
    $\langle Z_1 Z_{120}\rangle$ at $t=2.0$.
    Time denotes the QPU time for the quantum calculation and the wall-clock
    time for the classical calculations, in seconds.
    Speedup is the ratio of the corresponding classical runtime to the
     QPU time.}
    \label{tab:annulene_benchmark}
    \begin{tabular}{lccccccc}
    \toprule
    & \multicolumn{4}{c}{All observables (all $t$)}
    & \multicolumn{3}{c}{Boundary observable ($t=2.0$)} \\
    \cmidrule(lr){2-5}
    \cmidrule(lr){6-8}
    & Mean & Median & Time (s) & Speedup    & Error & Time (s) & Speedup \\
    \midrule
    Quantum  & 0.203 & 0.090 & 552  & --  & 0.044 & 59  & -- \\
    Majorana high  & -- & -- & 361,445  & 655$\times$  & -- & 40  & 0.68$\times$ \\
    Majorana low  & 0.014 & 0.014 & 75,113  & 136$\times$  & 0.001 & 10  & 0.17$\times$ \\
    TDVP ($\chi=16$)  & 0.020 & 0.020 & 2,477  & 4.49$\times$  & 0.018 & 2,477  & 42.0$\times$ \\
    TDVP ($\chi=8$)  & 0.025 & 0.023 & 888  & 1.61$\times$  & 0.094 & 888  & 15.1$\times$ \\

    \bottomrule
    \end{tabular}
\end{table}

Finally, we discuss break-even situations in which a strategy that uses quantum computation may be reasonable.
To this end, we considered a hybrid strategy in which only observables associated with low-fidelity hardware qubits were replaced by values obtained from Majorana propagation.
The hardware qubits were ranked according to 
\begin{equation}
    \begin{aligned}
S_q =&
-\log\!\left[1-\epsilon_{\rm ro}(p(q))\right]
+n_{\rm SX}(p(q))\{-\log[1-\epsilon_{\rm SX}(p(q))]\} \\
&+\frac{1}{2}\sum_r n_{\rm CZ}(p(q),r)
\{-\log[1-\epsilon_{\rm CZ}(p(q),r)]\}    
    \end{aligned}
\end{equation}
where \(p(q)\) is the index of the physical qubit assigned to logical qubit \(q\).
Here, \(\epsilon_{\rm ro}\), \(\epsilon_{\rm SX}\), and
\(\epsilon_{\rm CZ}\) are the historical readout, SX-gate, and CZ-gate
errors, respectively, \(n_{\rm SX}\) and \(n_{\rm CZ}\) denote the numbers of
occurrences of the corresponding gates in the transpiled circuit, and \(r\) denotes the physical qubit coupled to \(p(q)\) by a CZ-gate. Logical qubits were selected in
descending order of \(S_q\).
Starting from the highest-ranked qubits, we selected qubits sequentially and replaced all one- and two-qubit \(Z\) observables containing the selected qubits with the corresponding Majorana-propagation estimates, while adding the associated classical computational cost.
Majorana propagation is more amenable to near-ideal parallelization than TDVP, since the Majorana-propagation calculations for different observables are independent; we assumed ideal parallelization over 100 compute nodes for the Majorana calculations.
A similar hybrid strategy can also be constructed using TDVP.
For TDVP, the qubits were ranked according to the mean difference between observable estimates at successive bond dimensions over the observables containing each qubit.

Figure~\ref{fig:annulene_120q}c shows the resulting trade-off between accuracy and computational time for calculating all 7,260 observables at \(t=2.0\).
The hybrid Quantum--Majorana calculation (red dashed curve) achieved a mean absolute error below \(0.1\) in approximately \(449\)~s.
At this point, the hybrid calculation retained the QPU estimates for the 83 higher-fidelity logical qubits, while Majorana propagation was used selectively for 3,774 observables involving the 37 lowest-fidelity logical qubits.
For comparison, the fastest TDVP--Majorana result satisfying the same accuracy criterion, obtained with \(\chi=2\) (blue curve), required approximately \(534\)~s, whereas the standalone Majorana and TDVP calculations required approximately \(751\) and \(888\)~s, respectively.
In addition, Supplementary Note~4 shows a further improvement of the hybrid strategy in some time-accuracy regions by using advanced error mitigation and suppression. 
Thus, for a given parallelization model and accuracy target, there can exist a regime in which combining QPU results with selectively computed classical observables provides a favorable time-to-solution.

The present quantum--classical crossover is unlikely to remain fixed. More sophisticated
classical approaches, including transverse tensor-network contraction and
belief-propagation-based tensor-network methods, can substantially change the
comparison for particular geometries and dynamics~\cite{Ouyang2026-ni, Tindall2026-qg}.
In addition, classical runtimes may be further reduced through improvements in the software stack, including more efficient intra-node and, where applicable, inter-node parallelization, as well as highly optimized code implementations. 
Conversely, the software stack can also tilt the balance in favor of the quantum side, e.g., sophisticated error mitigations and suppressions. 
One route from quantum--classical break-even towards quantum advantage is to move
to interaction structures that are more costly for classical algorithms, that is, two- and three-dimensional geometries. 
Two-dimensional connectivity is already available on superconducting processors, while platforms with more flexible connectivity include trapped-ion and neutral-atom processors~\cite{Kim2023-xg,Brown2016-be,Maskara2025-fa}. The locality-based
compression workflow is not intrinsically restricted to one-dimensional open
chains and can in principle be extended to any geometries~\cite{D-Anna2025-wk}.
A further direction for applications is to remove the translational-symmetry assumption and apply the general local Hilbert--Schmidt objective (described in Supplementary Note~1) to inhomogeneous chemical systems. 
Such extensions may reach regimes that are intractable for full-system classical dynamics, while the local circuit optimization may itself require high-performance computing resources, including massively parallel CPU or distributed GPU environments.


\section*{Methods}

\subsection*{Hamiltonian construction}
We constructed effective $\pi$-electron Hamiltonians for benzene
(C$_6$H$_6$) and [60]annulene (C$_{60}$H$_{60}$). 
Four controlled
approximations were introduced in constructing these models:
(i) restriction of the electronic degrees of freedom to the carbon
$2p_z$-derived $\pi$ active space,
(ii) enforcement of exact point-group and translational periodicity,
(iii) truncation of interactions with localized orbitals five or more positions apart, and
(iv) omission of weak interactions of three- and four-centre terms. Here, the number of centres refers to the number of distinct orbitals, not spin orbitals.
Note that the use of chemical models with fewer approximations will be important for end-to-end chemical calculations on quantum devices in future studies. This will require, for example, the design of workflows that generate circuits with high representational power.

We used neutral singlet reference states for both molecules. For benzene, an
ideal $D_{\rm{6h}}$-symmetric geometry was used, and a conventional all-electron
restricted Hartree--Fock (RHF) calculation was performed with the cc-pVDZ
basis set. For [60]annulene, the molecular geometry was optimized within the
exact $D_{\rm{12h}}$-symmetric subspace, starting from a PM7 preoptimization~\cite{Stewart2013-tu}, followed by PBE/3-21G and BLYP45/def2-SVP refinements~\cite{Perdew1996-qc, Becke1988-sd, Lee1988-fj, Stawski2024-ho}.
We performed an ensemble-RHF calculation with the cc-pVDZ basis set to preserve $D_{12h}$ symmetry. The degenerate frontier orbital pair was assigned occupations of one electron each, while the other orbitals retained the conventional closed-shell occupations. The initial density matrix was averaged over the 12 equivalent $30^\circ$ rotations.

Approximation (i) was introduced by restricting the subsequent Hamiltonian to
the $\pi$ active space. The active orbitals were selected using the atomic
valence active space (AVAS) procedure~\cite{Sayfutyarova2017-fr} with carbon $2p_z$ atomic orbitals as
targets. STO-3G was used only as the minimal atomic-orbital reference for the
AVAS projection, whereas the molecular orbitals entering the electronic
structure calculation were obtained with cc-pVDZ. An AVAS threshold of 0.2
selected six $\pi$-type spatial orbitals containing six active electrons for
benzene and 60 $\pi$-type spatial orbitals containing 60 active electrons for
[60]annulene, yielding the complete active spaces CAS(6e,6o) and CAS(60e,60o),
respectively. The selected benzene orbitals were localized using the Boys
procedure~\cite{Foster1960-kw}, whereas the [60]annulene orbitals were localized using the
Pipek--Mezey procedure~\cite{Pipek1989-up}. In both cases, the localized orbitals were assigned to
individual carbon atoms, ordered around the molecular ring according to their
polar angles, and given a consistent orbital-phase convention.

The effective one-electron integrals were constructed by incorporating the contribution from the doubly occupied inactive orbitals into the active-space Hamiltonian. The two-electron integrals were obtained by an exact four-index atomic-orbital-to-molecular-orbital transformation within the selected active space. 

Approximation (ii) was introduced by enforcing exact translational periodicity of the effective Hamiltonian. The one- and two-electron integrals were averaged over symmetry-equivalent translations around the ring. Because the molecular geometries were already constrained to the corresponding point-group symmetries, this averaging produced only negligibly small changes in the integrals and mainly removed residual numerical symmetry breaking.

Approximation (iii) restricts the spatial range of
the interactions. Let $N$ denote the total number of qubits and $p,q,r,s \in \{1,\ldots,N/2\}$ denote the localized
spatial orbitals. For the periodic system,
we define the circular orbital distance as
\begin{equation}
d(p,q)
=
\min\left(
|p-q|,
\frac{N}{2}-|p-q|
\right).
\end{equation}
A one-electron integral $h_{pq}$ was retained only when
\begin{equation}
d(p,q)\leq 3,
\end{equation}
and a two-electron integral $(pq|rs)$ was retained only when
\begin{equation}
\max_{a,b\in\{p,q,r,s\}} d(a,b)\leq 3.
\label{eq:interaction_range}
\end{equation}
Thus, the distance criterion can retain interactions involving as many as four
successive localized spatial orbitals, while longer-range couplings are set to
zero.

Approximation (iv) is related to boundary terms under the Jordan--Wigner transformation. 
For example, the hopping term can be represented as 
\begin{equation}
\hat a_{l}^{\dagger}\hat a_{m}
+ \hat a_m^{\dagger}\hat a_l,
\end{equation}
where $\hat a_{l}^{\dagger}$ and
$\hat a_{l}$ are fermionic creation and annihilation operators for
spin-orbital index $l$, with $l<m$ in the following expressions.
Under the interleaved order in this study, the Jordan--Wigner mapping contains a
string of the form
\begin{equation}
\hat a_l^{\dagger}\hat a_m+\hat a_m^{\dagger}\hat a_l
\mapsto
\frac12\left(
X_l Z_{l+1}\cdots Z_{m-1}X_m+
Y_l Z_{l+1}\cdots Z_{m-1}Y_m\right).
\label{eq:jw_hopping}
\end{equation}
For bulk terms, the hopping range is limited by $d(p,q)$ and $d(a,b)$, and the maximum qubit locality for the hopping term is seven.
On the other hand, for hopping across the periodic cut (e.g., $l=1$ and $m=N -1$), this representation can contain an almost
system-wide $Z$ string.

Such strings can be shortened by introducing a particle-number parity operator~\cite{Lieb1961-pk,Chen2026-rb}. 
The total fermion-parity operator is
\begin{equation}
\hat P=(-1)^{\hat N_{\rm{e}}}
=
\prod_{k=1}^{N} Z_k,
\label{eq:parity}
\end{equation}
where $\hat N_{\rm{e}}$ denotes the number of electrons in the active space, $\hat N_{\rm{e}} = \sum_{l=1}^{N} \frac{1-Z_l}{2}$.
Multiplying the two Pauli
components in Eq.~\eqref{eq:jw_hopping} from the left by $\hat P$ gives
\begin{equation}
\begin{aligned}
\hat P(X_l Z_{l+1}\cdots Z_{m-1}X_m)
&=
-Z_1\cdots Z_{l-1}Y_l Y_m Z_{m+1}\cdots Z_N,
\end{aligned}
\label{eq:parity_components}
\end{equation}
and
\begin{equation}
\begin{aligned}
\hat P(Y_l Z_{l+1}\cdots Z_{m-1}Y_m)
&=
-Z_1\cdots Z_{l-1}X_l X_m Z_{m+1}\cdots Z_N,
\end{aligned}
\label{eq:parity_components2}
\end{equation}
where $Z_l X_l=i Y_l$ and $Z_l Y_l=-i X_l$. Since
$\hat P^2=I$, Eq.~\eqref{eq:jw_hopping} can be represented within a
fixed-parity sector with eigenvalue $p=1$ (even) or $-1$ (odd). By taking odd parity, it becomes
\begin{equation}
\begin{aligned}
\hat P^2(\hat a_l^{\dagger}\hat a_m+\hat a_m^{\dagger}\hat a_l)
\mapsto
&-\frac{p}{2}\left(
X_m Z_{m+1} \cdots Z_N Z_1 \cdots Z_{l-1} X_l + Y_m Z_{m+1} \cdots Z_N Z_1 \cdots Z_{l-1} Y_l\right)\\
=&\frac{1}{2}\left(
X_m Z_{m+1} \cdots Z_N Z_1 \cdots Z_{l-1} X_l + Y_m Z_{m+1} \cdots Z_N Z_1 \cdots Z_{l-1} Y_l\right)
\end{aligned}
\label{eq:parity_complement}
\end{equation}
which is the same representation as a bulk term on the ring topology.

Long-range \(Z\)-operator terms can also arise in three- and four-center terms.
However, the three- and four-center terms except for the density-assisted hopping terms cannot be treated in the same manner. For the two-center hopping and density-assisted hopping terms, the parity transformation reverses the coefficient sign once. For the other three- and four-center terms, the number of \(X\) and \(Y\) operators is doubled, so the corresponding sign factor is applied twice, resulting in a different coefficient sign. 
We verified this for benzene and found that the error in \(\langle Z_1\rangle\) introduced by removing these terms at \(t=2.0\) was negligibly small, on the order of \(10^{-5}\). We therefore removed these terms in the present calculations. This constitutes approximation (iv).
In Supplementary Note~2, we describe the treatment of all types of interactions included in the Hamiltonian, as well as prescriptions for incorporating all of them.

Electronic-structure calculations and active-space integral generation were performed using PySCF~\cite{Sun2020-sa}, with the PM7 preoptimization performed using MOPAC~\cite{Stewart2013-tu}. Fermionic Hamiltonians were transformed to qubit operators using OpenFermion~\cite{McClean2020-zp}, and quantum circuits and hardware experiments were implemented using Qiskit~\cite{Javadi-Abhari2024-qm} version 2.4.2.

\subsection*{Hardware settings}

Both benzene and annulene models were executed using the IBM Runtime Estimator with dynamical decoupling, TREX, and gate twirling enabled. Zero-noise
extrapolation was implemented through gate folding with noise factors of
$(1, 3, 5)$ and 50,000 shots per noise factor. The extrapolators
exponential, polynomial\_degree\_2, and linear were provided to the Runtime,
and the Runtime-selected zero-noise extrapolation estimate was used. C$_6$H$_6$ was executed on
ibm\_kawasaki with Qiskit optimization level 3 and a fixed layout, whereas
$D_{\rm{12h}}$ [60]annulene was executed on ibm\_boston with optimization level 3 and
a backend-aware automatic layout.

For C$_6$H$_6$ on \texttt{ibm\_kawasaki}, logical qubits
$(q_1,\ldots,q_{12})$ were mapped to physical qubits
(72, 73, 79, 93, 92, 91, 90, 89, 78, 69, 70, 71). For the 120-qubit [60]annulene
calculation on \texttt{ibm\_boston}, logical qubits $(q_1,\ldots,q_{120})$
were mapped to physical qubits
(3, 4, 5, 6, 7, 8, 9, 10, 11, 12, 13, 14, 15, 19, 35, 34, 33, 32, 31, 30, 29, 28, 27, 26, 25, 37, 45, 46, 47, 48, 49, 50, 51, 52, 53, 54, 55, 59, 75, 74, 73, 79, 93, 94, 95, 99, 115, 114, 113, 119, 133, 134, 135, 139, 155, 154, 153, 152, 151, 138, 131, 130, 129, 118, 109, 110, 111, 98, 91, 90, 89, 78, 69, 68, 67, 66, 65, 77, 85, 86, 87, 97, 107, 106, 105, 117, 125, 126, 127, 137, 147, 146, 145, 144, 143, 136, 123, 122, 121, 116, 101, 102, 103, 96, 83, 82, 81, 76, 61, 62, 63, 56, 43, 42, 41, 36, 21, 22, 23, 16).
The initial state of [60]annulene is $|1001$ $1001$ $1001$ $1001$ $1001$ $1001$ $1001$ $1001$ $1001$ $1001$ $0101$ $1001$ $1001$ $1001$ $1001$ $1001$ $1001$ $1001$ $1001$ $1010$ $1001$ $1001$ $1001$ $1001$ $1001$ $1001$ $1001$ $1001$ $1001$ $1000\rangle$.

\subsection*{Majorana-propagation dynamics}

Classical reference dynamics were calculated with the \texttt{monoprop}
package using Majorana propagation~\cite{Miller2025-bm}. In this
approach, observables are expanded in a Majorana-operator basis and evolved in
the Heisenberg picture through the time-evolution circuit, with small operator
coefficients truncated during propagation. Time evolution was represented with
a second-order Trotter decomposition.

We choose the high-accuracy setting of $\Delta t=0.05$ and $\epsilon_{\mathrm{cut}}=10^{-7}$ as the reference for observable values, and the low-accuracy setting of $\Delta t=0.2$ and $\epsilon_{\mathrm{cut}}=10^{-3}$ as the runtime reference.
Relaxing
$\epsilon_{\mathrm{cut}}$ further did not produce a commensurate reduction in
runtime, suggesting that the benchmark was limited substantially by
system-size-dependent overhead rather than by Majorana-string propagation
alone. We therefore did not use a looser cutoff to define the classical baseline.

Both the Majorana-propagation and TDVP (next section) calculations were performed on a single compute node with 36 CPU cores [Intel Xeon Gold 6254 (3.10 GHz)].

\subsection*{Time-dependent variational principle for matrix-product states}
The classical tensor-network calculations were performed using the two-site time-dependent variational principle implemented in ITensor~\cite{Fishman2022-et}. The 120-qubit periodic Pauli Hamiltonian was converted into a matrix-product operator (MPO), while the wave function was represented as an open-boundary matrix-product state (MPS); thus, the periodic interaction terms were retained in the Hamiltonian even though the tensor network itself had an open-boundary topology. The MPS was initialized as the same antiferromagnetic product state used in the quantum-device calculations. Real-time evolution was carried out up to \(t=2.0\) with a time step of \(\Delta t=0.05\). We used the two-site TDVP algorithm with state normalization after each time step, no singular-value cutoff (\(\texttt{cutoff}=0\)), and maximum bond dimensions \(\chi=1,2,4,8,16,\) and \(32\). At every time step, all 120 single-qubit \(Z_l\) observables and all \(7{,}140\) two-qubit \(Z_lZ_m\) observables were evaluated, giving \(7{,}260\) observables in total. The reported TDVP computation times include both the time evolution and the evaluation of these observables, with the time evolution dominating the computational time.

\section*{Data availability}
The data supporting the findings of this study are available in the associated
GitHub repository at \url{https://github.com/sk888ks/macrocycle_dynamics_open.git}.

\section*{Code availability}
The code used to generate and analyse the results reported in this study is
available in the associated GitHub repository at
\url{https://github.com/sk888ks/macrocycle_dynamics_open.git}.

\section*{Acknowledgements}
A part of this work was performed for Council for Science, Technology and Innovation (CSTI), Cross-ministerial Strategic Innovation Promotion Program (SIP), “Promoting the application of advanced quantum technology platforms to social issues” (Funding agency: QST).
Also, a part of this work was supported by MEXT Quantum Leap Flagship Program Grants No. JP- MXS0118067285 and No. JPMXS0120319794. 
Part of the calculations was performed on the Mitsubishi Chemical Corporation (MCC) HPC system “NAYUTA”, where “NAYUTA” is a nickname for MCC HPC and is not a product or service name of MCC.
We acknowledge the use of IBM Quantum services for experiments in this paper. The views expressed are those of the authors, and do not reflect the official policy or position of IBM or the IBM Quantum team.
All or part of the results in this paper were obtained using the ABCI-Q system provided by G-QuAT, AIST.
S.K. thanks Nathan Earnest and Yuri Kobayashi for technical discussions of error mitigation and Takao Kobayashi for advice on molecule selection.
S.K. thanks Q-CTRL for technical support with Fire Opal. 

\section*{Author contributions}
S.K. conceived the study, developed the methodology, performed the calculations
and quantum-hardware experiments, analysed the data, and wrote the manuscript.
K.S., R.S., H.N. and Q.G. discussed the chemical modelling and construction of
the electronic Hamiltonians with S.K. Toshinari Itoko and Takashi Imamichi contributed to calibration and
execution of the quantum-hardware experiments and discussed the results with
S.K. N.Y. discussed the quantum algorithms and provided methodological advice
to S.K. All authors discussed the results and reviewed the manuscript.

\section*{Competing interests}
The authors declare no competing interests.

\putbib[pc]
\end{bibunit}

\clearpage

\makeatletter
\let\title\savedtitle
\let\author\savedauthor
\let\date\saveddate
\let\maketitle\savedmaketitle
\let\@maketitle\savedatmaketitle
\makeatother
\title{\textbf{Supplementary Note for\\Quantum--classical break-even in electronic dynamics of macrocyclic molecules}}
\author{\parbox{0.94\textwidth}{\centering
Shu Kanno$^{1,2}$, Kenji Sugisaki$^{3,2,4}$, Takashi Imamichi$^{5}$, Toshinari Itoko$^{5,2}$, \\ Rei Sakuma$^{6,2}$, Hajime Nakamura$^{2}$, Qi Gao$^{1,2}$ and Naoki Yamamoto$^{2,7}$\\[0.6em]
\small $^{1}$Mitsubishi Chemical Corporation, Science \& Innovation Center, Yokohama 227-8502, Japan\\
\small $^{2}$Quantum Computing Center, Keio University, 3-14-1 Hiyoshi, Kohoku-ku, Yokohama 223-8522, Japan\\
\small $^{3}$Deloitte Tohmatsu LLC, 3-2-3 Marunouchi, Chiyoda-ku, Tokyo 100-8363, Japan\\
\small $^{4}$Centre for Quantum Engineering, Research and Education, TCG Centres for Research and Education in Science and Technology, Sector V, Salt Lake, Kolkata 700091, India\\
\small $^{5}$IBM Research, 19-21 Nihonbashi Hakozaki-cho, Chuo-ku, Tokyo 103-8510, Japan\\
\small $^{6}$Materials Informatics Initiative, RD Technology \& Digital Transformation Center, JSR Corporation, 3-103-9 Tonomachi, Kawasaki-ku, Kawasaki 210-0821, Japan\\
\small $^{7}$Department of Applied Physics and Physico-Informatics, Keio University, 3-14-1 Hiyoshi, Kohoku-ku, Yokohama 223-8522, Japan}}
\date{}
\maketitle
\begin{bibunit}[naturemag]
\setcounter{figure}{0}
\renewcommand{\figurename}{Supplementary Figure}
\renewcommand{\thefigure}{\arabic{figure}}

\setcounter{table}{0}
\renewcommand{\tablename}{Supplementary Table}
\renewcommand{\thetable}{\arabic{table}}

\setcounter{section}{0}
\renewcommand{\thesection}{Supplementary Note \arabic{section}}

\setcounter{equation}{0}
\renewcommand{\theequation}{\arabic{equation}}

\section{Local Hilbert--Schmidt objective}

\subsection{Theory}

Circuit compression approximates a target unitary operator with a shallower
variational circuit. For a full $N$-qubit system, let $V_{\mathrm{ref}}$ denote
the reference evolution and $U_{\mathrm{evol}}$ the variational circuit. For
each corresponding qubit pair in the doubled register, we define
\begin{equation}
|\Phi_j^+\rangle=\frac{|00\rangle_j+|11\rangle_j}{\sqrt{2}},
\end{equation}
\begin{equation}
\Pi_j=|\Phi_j^+\rangle\langle\Phi_j^+|.    
\end{equation}

For a $K$-qubit register, let
\begin{equation}
|\Phi_K\rangle=\bigotimes_{k=1}^{K}|\Phi_k^+\rangle,
\end{equation}
and
\begin{equation}
    |\Psi_{U_{\mathrm{evol}},V_{\mathrm{ref}}}\rangle
=
(U_{\mathrm{evol}}\otimes V_{\mathrm{ref}}^*)|\Phi_N\rangle,
\end{equation}
where $^*$ denotes complex conjugation. For any
operator $O$ on the doubled register, we use the notation
\begin{equation}
\langle O\rangle_{U_{\mathrm{evol}},V_{\mathrm{ref}}}
\equiv
\langle\Psi_{U_{\mathrm{evol}},V_{\mathrm{ref}}}|
O
|\Psi_{U_{\mathrm{evol}},V_{\mathrm{ref}}}\rangle.    
\end{equation}

\begin{figure}[t]
    \centering
    \includegraphics[width=1.0\linewidth]{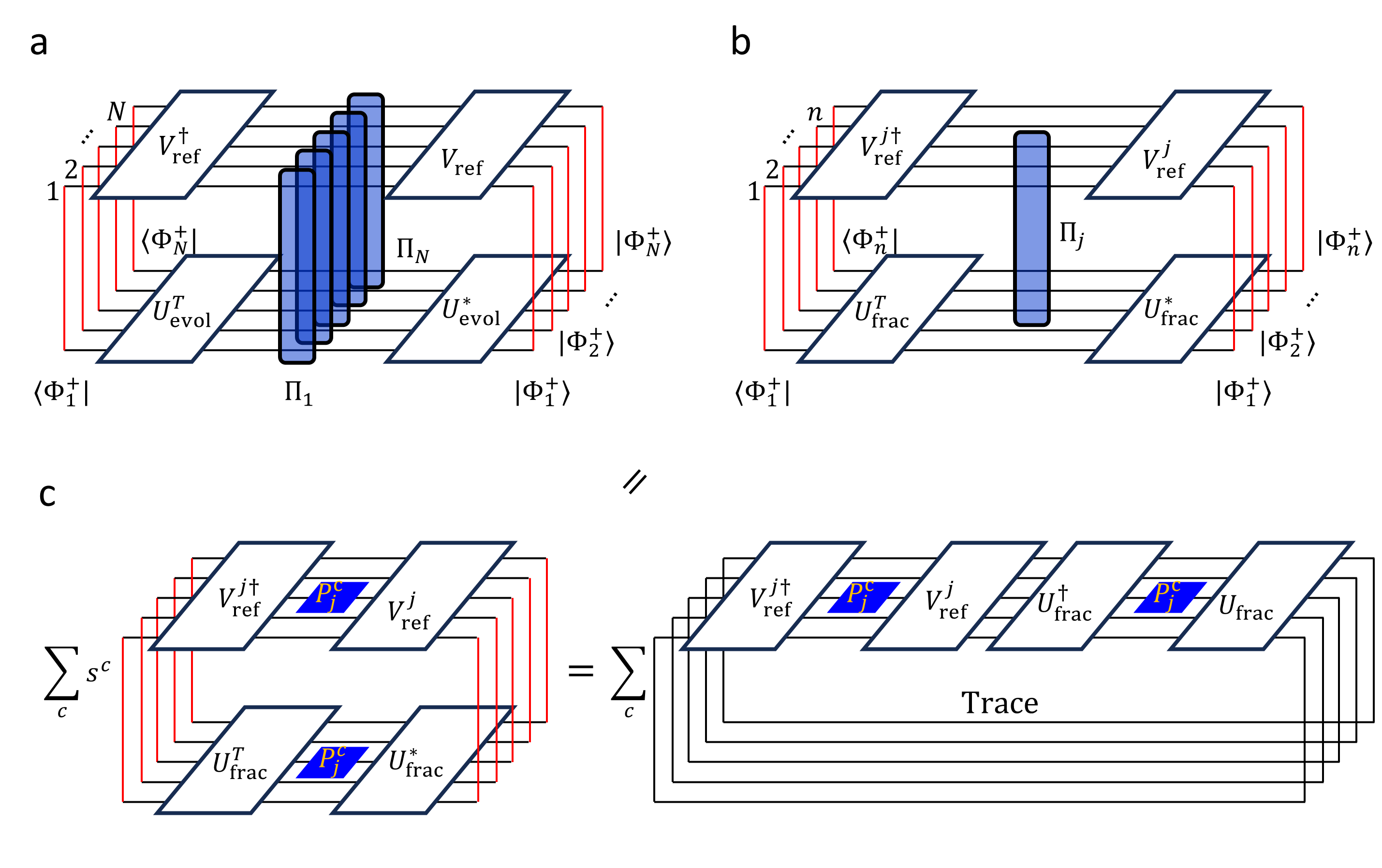}      
    \caption{Overlaps of HST and LHST. (a) HST representation for an $N$-qubit system. (b) LHST conventional representation for an $n$-qubit fractional system. $P_j^c \in \{I,X_j, Y_j,Z_j\}$, and $s^c$ takes $-1$ if $P_j^c = Y_j$ and $+1$ otherwise. (c) LHST representation with channel--state duality for an $n$-qubit fractional system.}
    \label{si:fig:s_lhst}
\end{figure}

The objective of the normalized Hilbert--Schmidt test for the full system is
\begin{equation}
C_{\rm{HST}} = 1-F_{\mathrm{HS}}, \label{si:eq:s_hst}
\end{equation}
\begin{equation}
F_{\mathrm{HS}}
=\frac{1}{4^N}
\left|
\Tr\!\left(V_{\mathrm{ref}}^\dagger U_{\mathrm{evol}}\right)
\right|^2
=
\left\langle
\prod_{j=1}^{N}\Pi_j
\right\rangle_{U_{\mathrm{evol}},V_{\mathrm{ref}}},
\label{si:eq:s_hst2}
\end{equation}
which is the probability of a global Bell projection over all corresponding
qubit pairs (Supplementary Figure~\ref{si:fig:s_lhst}a).

The local Hilbert--Schmidt test replaces the global Bell projection by local
Bell projections. In the general case, the objective averages the local costs
over all system qubits,
\begin{equation}
C_{\mathrm{LHST}}
=\frac{1}{N}\sum_{j=1}^{N}\Tilde{C}_{\mathrm{LHST}}^{(j)},
\qquad
\Tilde{C}_{\mathrm{LHST}}^{(j)}
=
1-\langle\Pi_j\rangle_{U_{\mathrm{evol}},V_{\mathrm{ref}}}.
\label{si:eq:s_lhst_general}
\end{equation}
The HST and LHST costs satisfy the bounds
\begin{equation}
C_{\mathrm{LHST}}
\le
C_{\mathrm{HST}}
\le
N C_{\mathrm{LHST}}.
\end{equation}
Thus, the two cost functions vanish simultaneously when the target and variational unitaries coincide up to a global phase~\cite{Khatri2019-wb}.

As shown by a previous study~\cite{Mizuta2022-ql}, the evaluation of the local cost \(\Tilde{C}_{\mathrm{LHST}}^{(j)}\) can be restricted to a subsystem containing the approximate causal cone of the short-time evolution around qubit \(j\) and the relevant causal cone of the ansatz. Therefore, instead of treating the full \(N\)-qubit system, it is sufficient to consider an \(n\)-qubit subsystem with \(n < N\). 
We define the LHST for the fractional system (Supplementary Figure~\ref{si:fig:s_lhst}b) using the $n$-qubit reference $V_{\mathrm{ref}}^j$ and circuit $U_{\mathrm{frac}}$ as
\begin{equation}
C_{\mathrm{LHST}}^{\rm{frac}}
=\frac{1}{N}\sum_{j=1}^{N}C_{\mathrm{LHST}}^{(j)},
\end{equation}
\begin{equation}
    C_{\mathrm{LHST}}^{(j)}
= 1-\langle\Pi_j\rangle_{U_{\mathrm{frac}},V_{\mathrm{ref}}^j},
\end{equation}
\begin{equation}
    |\Psi_{U_{\mathrm{frac}},V_{\mathrm{ref}}^j}\rangle
=
(U_{\mathrm{frac}}\otimes (V_{\mathrm{ref}}^j)^*)|\Phi_n\rangle,
\end{equation}
and
\begin{equation}
\langle O\rangle_{U_{\mathrm{frac}},V_{\mathrm{ref}}^j}
\equiv
\langle\Psi_{U_{\mathrm{frac}},V_{\mathrm{ref}}^j}|
O
|\Psi_{U_{\mathrm{frac}},V_{\mathrm{ref}}^j}\rangle.    
\end{equation}
That is, $\langle O\rangle_{U_{\mathrm{frac}},V_{\mathrm{ref}}^j}
$ is an evaluation of $\langle O\rangle_{U_{\mathrm{evol}},V_{\mathrm{ref}}}$ within the $n$-qubit subsystem.

For a translationally symmetric
Hamiltonian and a variational circuit with the same symmetry, symmetry-equivalent sites have identical local costs, so
the average in $C_{\mathrm{LHST}}^{\mathrm{frac}}$ can be restricted to a single unit cell,
\begin{equation}
C_{\mathrm{LHST}}^{\mathrm{per}}
=
\frac{1}{N_{\mathrm{uc}}}
\sum_{j\in\mathcal S_{\mathrm{uc}}}C_{\mathrm{LHST}}^{(j)},
\label{si:eq:s_lhst_periodic}
\end{equation}
where $N_{\mathrm{uc}}=|\mathcal S_{\mathrm{uc}}|,
$ and $\mathcal S_{\mathrm{uc}}$ is the set of qubits in one unit cell.

A direct evaluation of the LHST on this subsystem requires a doubled \(2n\)-qubit system because of the Bell-state construction. 
Here, we show that the same local cost can be deterministically evaluated using only an \(n\)-qubit tensor network by exploiting channel--state duality.
Expanding the Bell-state projector
in single-qubit Pauli operators gives
\begin{equation}
\Pi_j
=
\frac14\left(
I\otimes I
+X_j\otimes X_j
-Y_j\otimes Y_j
+Z_j\otimes Z_j
\right),
\label{si:eq:s_bell_proj}
\end{equation}
where $X_j$, $Y_j$, and $Z_j$ are the Pauli operators acting on the target
qubit $j$. Therefore,
\begin{equation}
\langle\Pi_j\rangle_{U_{\mathrm{frac}},V_{\mathrm{ref}}^j}
=
\frac14\left[
1
+\langle X_j\otimes X_j\rangle_{U_{\mathrm{frac}},V_{\mathrm{ref}}^j}
-\langle Y_j\otimes Y_j\rangle_{U_{\mathrm{frac}},V_{\mathrm{ref}}^j}
+\langle Z_j\otimes Z_j\rangle_{U_{\mathrm{frac}},V_{\mathrm{ref}}^j}
\right].
\label{si:eq:s_bell_pauli}
\end{equation}

Then we use channel--state duality for the
maximally entangled state,
\begin{equation}
\langle\Phi_n|A\otimes B|\Phi_n\rangle
=
\frac{1}{2^n}\Tr(AB^{\mathsf T}),
\label{si:eq:s_duality}
\end{equation}
where $A$ and $B$ are $n$-qubit operators and ${\mathsf T}$ denotes transpose.
Applying Eq.~\eqref{si:eq:s_duality} to the Pauli-expanded Bell projector yields
the single-register form
\begin{equation}
C_{\mathrm{LHST}}^{(j)}
=
1-\frac14\left[
1+\frac{1}{2^n}
\sum_{P_j\in\{X_j,Y_j,Z_j\}}
\Tr\!\left(
U_{\mathrm{frac}}^\dagger P_j U_{\mathrm{frac}}
V_{\mathrm{ref}}^{j\dagger} P_j V_{\mathrm{ref}}^j
\right)
\right],
\label{si:eq:s_lhst_trace}
\end{equation}
where we used the relation $Y^{\mathsf T}=-Y$. Equation~\eqref{si:eq:s_lhst_trace} can therefore be
evaluated on the original $n$-qubit tensor network without explicitly
constructing the doubled system. 
Note that alternative Monte-Carlo sampling
estimators can avoid an explicit doubled-register contraction, while incurring additional sampling overhead~\cite{Mizuta2022-ql}.

\subsection{Calculation conditions}

For each local target cost, we prepare an accurate matrix-product-operator (MPO)
representation of the short-time ($\Delta t$) reference evolution. 
We first prepare the fractional $n$-qubit Hamiltonian around the $j$-th qubit, $H_{\rm{frac}}^j$. 
In this study, we choose $n=9$ and the Hamiltonian qubit indices $j-4, j-3, \dots, j,\dots, j+4$ for $j=5, 6$ in benzene; $n=17$ and indices $j-8, j-7, \dots, j,\dots, j+8$ for $j=11, 12, \dots, 20$ in [60]annulene. 
Then, the reference propagator is
constructed with a second-order Trotter formula,
\begin{equation}
    V_{\mathrm{ref}}^j
    =S_2\!\left(\delta t; H_{\rm{frac}}^j \right)^{m_{\mathrm T}},
    \label{si:eq:reference_trotter}
\end{equation}
where $S_2(\delta t; H_{\rm{frac}}^j)$ denotes one second-order Trotter step of $H_{\rm{frac}}^j$ with duration $\delta t$, and $\Delta t =\delta t \times m_{\rm{T}}$. We choose $\delta t = 10^{-3}$ and $m_{\rm{T}} = 200$ for $\Delta t = 0.2$.

Each Pauli-rotation factor appearing within a Trotter step is constructed
directly as an MPO. For a Pauli string $P$ and the
substep duration $\tau$ assigned by the second-order Trotter formula, we consider the following decomposition,
\begin{equation}
    e^{-\iu c \tau P}
    =\cos(c\tau)I
    -i\sin(c\tau)P,
    \label{si:eq:pauli_rotation_mpo}
\end{equation}
where $c$ is a coefficient for the qubit Hamiltonian.
Both $I$ and $P$ can take an MPO representation with bond
dimension one, so their sum has bond dimension at most two before subsequent
MPO multiplications and truncations. 
This implementation avoids the direct calculation of $e^{-\iu c \tau P}$, which would be heavy if the qubit locality of $P$ is large.

The variational circuit $U_{\mathrm{frac}}$ is a brick-wall circuit of general
two-qubit $SU(4)$ unitaries. We minimize the local Hilbert--Schmidt objective in Eq.~\eqref{si:eq:s_lhst_trace}
with the limited-memory Broyden--Fletcher--Goldfarb--Shanno (L-BFGS) algorithm
for up to 128 iterations. The singular-value-decomposition cutoff in the
tensor-network calculations is $10^{-10}$. Tensor-network evaluations of the
objective and the associated circuit optimization were performed on NVIDIA
GPUs, including H100 accelerators.
Tensor-network calculations were performed using ITensor~\cite{Fishman2022-et}.

\begin{table}[t]
    \centering
    \footnotesize
    \caption{\textbf{Treatment of fermionic interaction terms.} Representative
    second-quantized operator forms are shown without numerical prefactors. Here
    $p,q,r,s$ denote spatial orbitals, $\sigma,\tau$ spin indices and
    $\bar\sigma$ the spin opposite to $\sigma$. ``Boundary handling'' denotes
    parity-based shortening of a Jordan--Wigner string across the periodic cut.
    A dash indicates that no odd-sector coefficient is required.}
    \label{si:tab:interaction_treatment}
    \setlength{\tabcolsep}{2pt}
    \renewcommand{\arraystretch}{1.12}

    \begin{tabularx}{\textwidth}{
        @{}
        >{\raggedright\arraybackslash\hsize=1.15\hsize}X
        >{\raggedright\arraybackslash\hsize=1.65\hsize}X
        >{\raggedright\arraybackslash\hsize=0.90\hsize}X
        >{\raggedright\arraybackslash\hsize=0.75\hsize}X
        >{\raggedright\arraybackslash\hsize=0.90\hsize}X
        >{\raggedright\arraybackslash\hsize=0.65\hsize}X
        @{}
    }
        \toprule
        Term type & Representative operator & Long-range $Z$ at periodic cut & Boundary handling & Odd-sector coefficient & Retained \\
        \midrule
        Constant & $I$ & No & No & -- & Yes \\
        On-site one-body & $\hat n_{p\sigma}$ & No & No & -- & Yes \\
        Ordinary hopping & $\hat a_{p\sigma}^{\dagger}\hat a_{q\sigma}+\mathrm{h.c.}$ & Yes & Yes & Same as bulk & Yes \\
        On-site Coulomb & $\hat n_{p\uparrow}\hat n_{p\downarrow}$ & No & No & -- & Yes \\
        Interorbital direct Coulomb & $\hat n_{p\sigma}\hat n_{q\tau}$ & No & No & -- & Yes \\
        Two-centre density-assisted hopping & $\hat n_{p\bar\sigma}(\hat a_{p\sigma}^{\dagger}\hat a_{q\sigma}+\mathrm{h.c.})$ & Yes & Yes & Same as bulk & Yes \\
        Three-centre density-assisted hopping & $\hat n_{r\tau}(\hat a_{p\sigma}^{\dagger}\hat a_{q\sigma}+\mathrm{h.c.})$ & Yes & Yes & Same as bulk & Yes \\
        Two-centre spin exchange & $\hat a_{p\uparrow}^{\dagger}\hat a_{q\downarrow}^{\dagger}\hat a_{p\downarrow}\hat a_{q\uparrow}+\mathrm{h.c.}$ & No (interleaved ordering) & No & -- & Yes \\
        Two-centre pair hopping & $\hat a_{p\uparrow}^{\dagger}\hat a_{p\downarrow}^{\dagger}\hat a_{q\downarrow}\hat a_{q\uparrow}+\mathrm{h.c.}$ & No (interleaved ordering) & No & -- & Yes \\
        Three-centre pair splitting/gathering & $\hat a_{p\sigma}^{\dagger}\hat a_{q\bar\sigma}^{\dagger}\hat a_{r\bar\sigma}\hat a_{r\sigma}+\mathrm{h.c.}$ & Yes & Yes & Sign reversed from bulk & No \\
        Three-centre correlated double hopping & $\hat a_{p\sigma}^{\dagger}\hat a_{q\bar\sigma}^{\dagger}\hat a_{r\bar\sigma}\hat a_{q\sigma}+\mathrm{h.c.}$ & Yes & Yes & Sign reversed from bulk & No \\
        Four-centre correlated double hopping & $\hat a_{p\sigma}^{\dagger}\hat a_{q\tau}^{\dagger}\hat a_{r\tau}\hat a_{s\sigma}+\mathrm{h.c.}$ & Yes & Yes & Sign reversed from bulk & No \\
        \bottomrule
    \end{tabularx}
\end{table}

\section{Treatment for the interactions}

The treatment of each type of interaction in the Hamiltonian is shown in Supplementary Table~\ref{si:tab:interaction_treatment}. 
For the one- and two-center interactions, except for hopping terms, the interleaved ordering employed in this work does not generate long-range \(Z\) operators in the boundary terms and does not break the rotational symmetry. The hopping terms can be treated by using the parity in the odd-electron sector, as described in the main text. For the three- and four-center interactions, long-range \(Z\) operators can in general appear, but the three-center density-assisted hopping terms, which have the largest contribution among these interactions, can be treated in the same manner as the hopping terms.
The neglected three- and four-center terms are expected to be small because the corresponding two-electron integrals involve products of spatially separated localized orbitals and are therefore strongly suppressed by their small spatial overlap. We actually checked that their effects are negligibly small for benzene.

The most general way to incorporate the interactions neglected in the present work is to evaluate the LHST in the form of Eq.~\eqref{si:eq:s_lhst_general}. 
Because this requires evaluating $\Tilde{C}_{\mathrm{LHST}}^{(j)}$ separately for $j \in \{1,2,\dots,N\}$, it increases the computational cost. Nevertheless, such an extension will likely be essential for treating inhomogeneous and multidimensional models in the future.

A simpler approach for incorporating the boundary terms has been proposed based on a fermionic ansatz~\cite{Kanasugi2023-yd}. By replicating the parameters rather than the gates themselves, the periodic-boundary treatment can be avoided. However, when we tested this approach using a particle-number-conserving brick-wall circuit, we obtained results far from the desired accuracy. 
To make this approach practical, a more expressive ansatz, such as the Hamiltonian-based variational ansatz~\cite{Wecker2015-fu} used in previous work~\cite{Kanasugi2023-yd}, would likely be required.

\begin{figure}[h]
    \centering
    \includegraphics[width=0.8\linewidth]{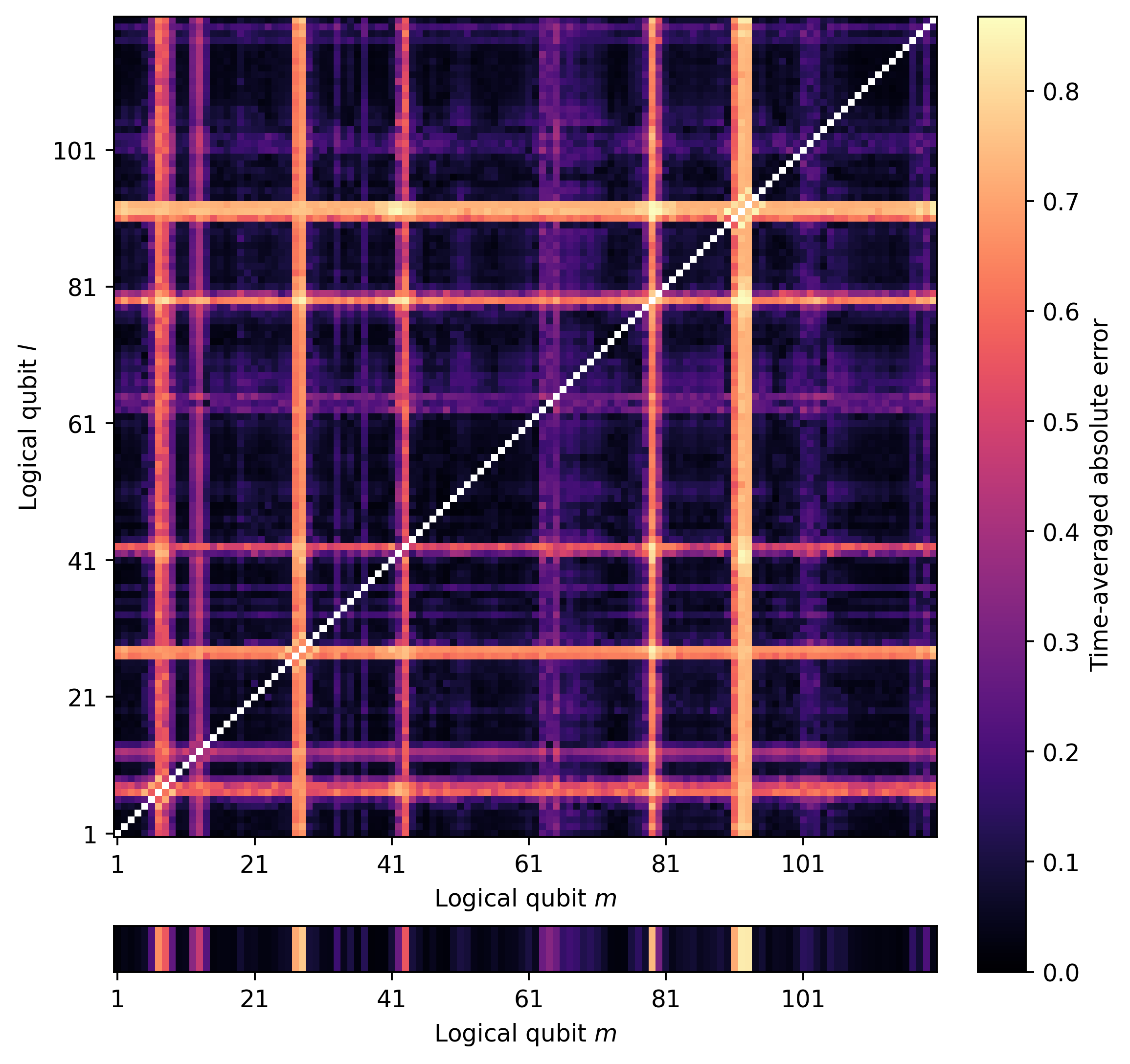}
    \caption{Spatial distribution of the errors in the measured 7,260 observables for [60]annulene. The upper 2D heatmap shows the absolute deviation of the 2-qubit correlators $\langle Z_m Z_l\rangle$ from the reference values, averaged over the ten time points, where each matrix element corresponds to a qubit pair $(m,l)$. The lower 1D heatmap shows the corresponding time-averaged absolute deviation of the 1-qubit expectation values $\langle Z_m\rangle$.}
    \label{si:fig:error_map}
\end{figure}

\section{Low-fidelity qubit evaluation}
One possible reason for the reduced accuracy and the difference between the median and mean values is that constructing a one-dimensional loop of 120 qubits on the device required the use of several low-fidelity qubits. 
Supplementary Figure~\ref{si:fig:error_map} shows a two-dimensional map of the errors in \(\langle Z_l Z_m\rangle\). The map indicates that large errors occur primarily when certain qubits are involved. 
Here, the two-qubit $Z$-observable signal-loss rate of a qubit is defined as the fraction of data points with a non-ZNE expectation-value magnitude of at most 0.05 among those with a high-accuracy Majorana-propagation reference magnitude of at least 0.5, considering all $Z_lZ_m$ observables involving that qubit over $t=0.2,0.4,\ldots,2.0$.
As a post hoc diagnostic, excluding all one- and two-qubit $Z$ observables involving any of the eight qubits whose signal-loss rate was at least 0.5, $(7,8,27,28,79,91,92,93)$, reduced the mean/median absolute errors of the ZNE results from 0.203/0.090 to 0.137/0.072, respectively.

\begin{figure}[h]
    \centering
    \includegraphics[width=0.8\linewidth]{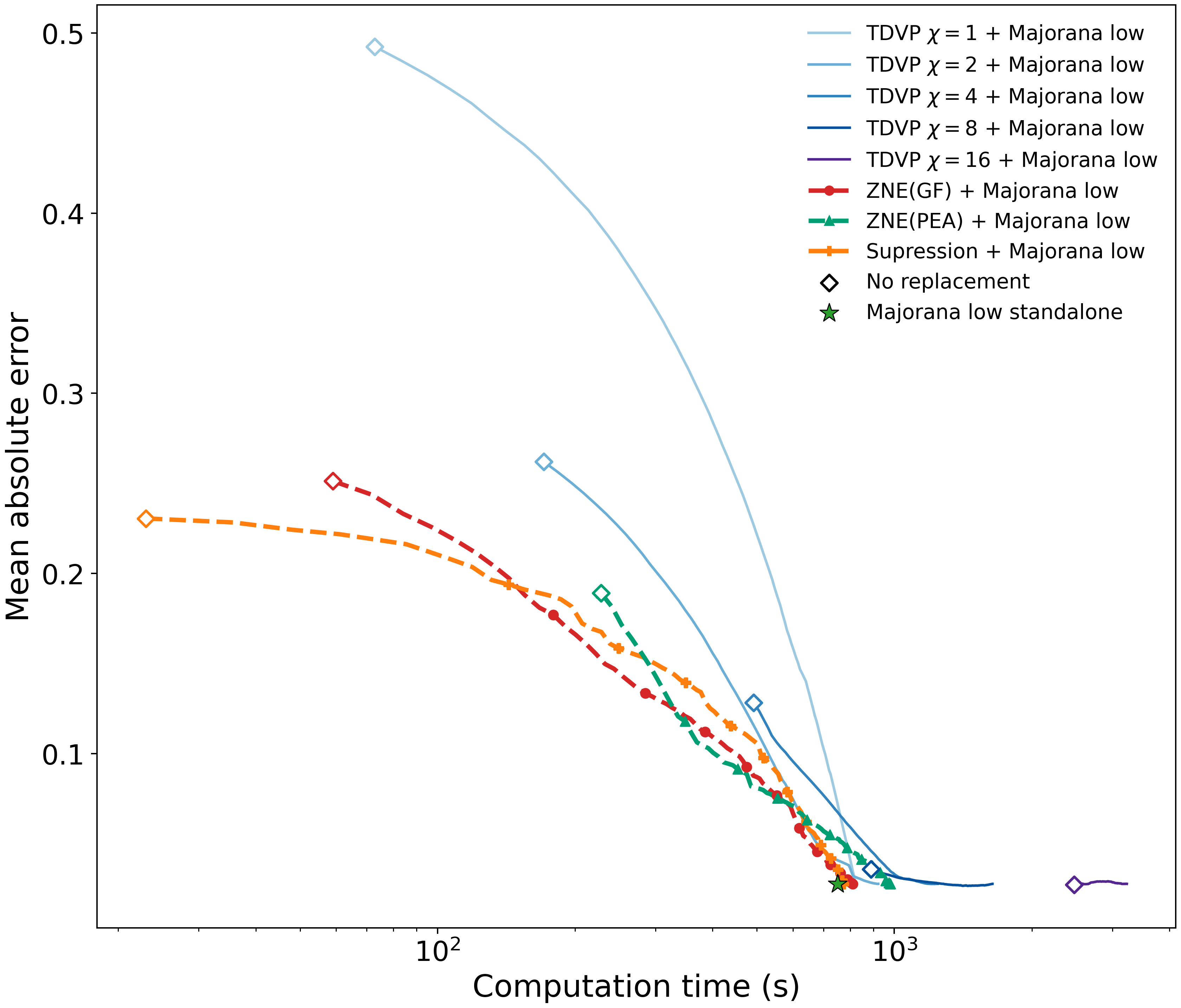}
    \caption{Time-to-accuracy trade-off at \(t=2.0\) for all 7,260 one- and two-qubit
\(Z\) observables of [60]annulene. The horizontal axis assumes ideal parallel execution of the Majorana propagation (low accuracy)
over 100 nodes.
Dashed curves represent results of QPU--Majorana hybrid strategies. 
The markers on the dashed curve indicate increments of ten selected logical qubits.}
\label{si:fig:hybrid_with_FO_PEA}
\end{figure}

\section{Evaluation for quantum implementations}
We investigate two advanced implementations of quantum-device execution: probabilistic error amplification (PEA) in ZNE by Qiskit~\cite{Javadi-Abhari2024-qm} and error suppression using Q-CTRL's Fire Opal~\cite{Mundada2023-rl}. PEA is a zero-noise extrapolation technique that probabilistically injects additional noise based on learned noise models. It enables fine control of noise amplification without relying on gate folding (GF). Fire Opal suppresses errors through automated circuit optimization and error-aware qubit mapping. ZNE was not applied in our experiments using Fire Opal.

For PEA, we used noise factors of \(1.0\), \(1.2\), and \(1.6\), with 2,000 randomizations and 64 shots per randomization. The twirling strategy was \texttt{active-circuit}. Layer-noise and readout-noise learning were performed using the Qiskit Runtime noise-learning procedure, and the physical-qubit layout from the GF experiment was reused. The remaining settings were the same as those used for the GF ZNE, including runtime-selected extrapolation from exponential, quadratic polynomial, and linear models. 

For Fire Opal, we used 50,000 shots per time step. Circuit optimization, qubit mapping, and error suppression were handled by Fire Opal. All qubits were measured in the computational basis, and the one- and two-qubit $Z$ expectation values were calculated locally from the returned samples.

Supplementary Figure~\ref{si:fig:hybrid_with_FO_PEA} compares the computational times under the same hybrid-analysis assumptions as in the main text, including ideal 100-node parallelization of Majorana propagation (low accuracy).  ``ZNE (GF) + Majorana'' (red dashed curve) corresponds to ``QPU + Majorana'' in the main text.
The PEA result (green dashed curve) is the fastest among the tested strategies at the target mean absolute error (MAE) $ \approx 0.1$: without replacement by Majorana-propagation results (hereafter Majorana replacement), its mean/median absolute errors are $0.189/0.112$, respectively, with a QPU time of $228$~s. At the first replacement point satisfying $\mathrm{MAE}\leq 0.1$, these errors decrease to $0.098$ and $0.053$, with a total hybrid time of $412$~s. This result demonstrates the importance of the error-mitigation configuration, although the larger shot budget used for PEA also contributes to the comparison. More advanced error-mitigation techniques used in an investigation of quantum advantage~\cite{Leviatan2026-fi, Aharonov2025-bp} may further improve the accuracy.

The error-suppressed result (yellow dashed curve) achieves a lower error than the ZNE(GF) without the Majorana replacement, despite not using explicit ZNE, and requires substantially less QPU time: its mean and median absolute errors are $0.230$ and $0.172$, respectively, with a QPU time of $23$~s. At the first replacement point satisfying $\mathrm{MAE}\leq 0.1$, the mean and median errors are $0.100/0.035$, with a total hybrid time of $510$~s. Thus, error suppression is advantageous for looser accuracy targets. On the other hand, more Majorana replacement is required to reach higher accuracy than with ZNE. Noise-induced attenuation of observable expectation values may contribute to this behavior. 
A technique called decay recovery, used in an investigation of quantum advantage~\cite{Hartnett2026-cq} to compensate for signal attenuation, may also be effective for pushing the quantum beneficial side.

These results suggest that more advanced quantum-execution and error-control techniques can improve the accuracy–runtime trade-off and leave further room for improvement as these techniques continue to advance.

\clearpage
\putbib[pc]
\end{bibunit}
\end{document}